\documentclass
[superscriptaddress,secnumarabic,amssymb,amsmath,nobibnotes,aps,prd,showkeys,showpacs,nofootinbib,onecolumn]{revtex4}%
\usepackage{graphicx}
\usepackage{epsf}
\usepackage{bm}
\usepackage{amsmath}
\usepackage{amsfonts}
\usepackage{amssymb}%
\providecommand{\U}[1]{\protect\rule{.1in}{.1in}}

\newcommand{\be}{\begin{equation}}
\newcommand{\ee}{\end{equation}}

\newcommand{\mincir}{\raise
-3.truept\hbox{\rlap{\hbox{$\sim$}}\raise4.truept\hbox{$<$}\ }}
\newcommand{\magcir}{\raise
-3.truept\hbox{\rlap{\hbox{$\sim$}}\raise4.truept\hbox{$>$}\ }}

\begin{document}
\title{New cosmological solutions in hybrid metric-Palatini gravity from dynamical symmetries}
\author{Andronikos Paliathanasis}
\email{anpaliat@phys.uoa.gr}
\affiliation{Institute of Systems Science, Durban University of Technology, Durban 4000,
Republic of South Africa}

\begin{abstract}
We investigate the existence of Liouville integrable cosmological models in
hybrid metric-Palatini theory. Specifically we use the symmetry conditions for
the existence of quadratic in the momentum conservation laws for the field
equations as constraint conditions for the determination of the unknown
functional form of the theory. The exact and analytic solutions of the
integrable systems which found in this study are presented in terms of
quadratics and Laurent expansions.

\end{abstract}
\date{\today}
\maketitle

\section{Introduction}

Modified theories of gravity \cite{mod1} are of special interest because they
preserve the geometric character of the gravitational theory, as it has been
introduced by General Relativity, while the modified gravitational theories
provide geometric mechanisms for the explanation of the recent cosmological
observations \cite{Teg,Kowal,Komatsu,planck,planck18}. One of the first
modification of the Einstein-Hilbert Action Integral which has been proposed
in the literature is that of quadratic gravity where all possible terms
quadratic in the curvature are introduced in the gravitational Action Integral
\cite{quad1}. Applications of quadratic gravity in cosmological studies can be
found in \cite{quad2,quad3,quad4,quad5} and references therein. When only the
quadratic term of the Ricciscalar is introduced into the gravitational Action
Integral, this specific model is also known as Starobinsky model for of
Inflation \cite{st1}. The importance of the latter gravitational model is that
it provides an inflationary scenario which is favoured by the observations.
Generalizations of the Starobinsky mode have been proposed in the literature,
where other functional forms of the Ricciscalar have been introduced in the
gravitational Action Integral \cite{st2,st3}, the proposed gravitational
models form a family of theories known as $f\left(  R\right)  $-theories of
gravity \cite{st4}, in which $R$ is the Ricciscalar corresponding to the
metric tensor.

Nowadays, there is a family of modified theories of gravity which are known as
$f\left(  X\right)  $-theories, where $X$ is a geometric invariant. In these
theories, a function $f$ of the geometric invariant $X$ is introduced into the
gravitational Action Integral in order to provide new geometrodynamical terms
which drive the dynamics in a way to explain the observations. In the case
where $X$ is the Ricciscalar of the underlying metric, then $f\left(
R\right)  $-theory is recovered \cite{st4}, while, when $X$ is the Palatini
curvature scalar $\mathcal{R}$, the affine $f\left(  \mathcal{R}\right)
$-theory is recovered \cite{st5}. Other $f-$theories of special interest are
the, $f\left(  T\right)  ~$teleparallel gravity
\cite{Ferraro,Ferraro06,Lin2010}, the $f\left(  G\right)  $ Gauss-Bonnet
theory \cite{gg1,gg2,gg3,gg4} and many others, for instance see
\cite{r1,r2,r3,r4,r5,r6} and references therein.

In this study we are interested in the integrability properties and in the
existence of exact and analytic cosmological solutions in the so-called hybrid
metric-Palatini theory \cite{m1,oo1,Capoz,Capoz2}. In this specific modified
theory a function $f$ of the Palatini curvature scalar $\mathcal{R}$ it is
introduced into the Einstein-Hilbert Action Integral. The resulting
gravitational theory is a fourth-order theory and equivalent with a scalar
tensor theory. There are various applications of hybrid metric Palatini theory
in gravitational physics, gravitational waves studied in \cite{hp1}, exact
wormhole solutions were studied in \cite{hp2,hp3}, while in
\cite{hp4,hp4a,hp4b} some exact and analytic cosmological solutions in the
classical and the quantum levels derived, while cosmological constraints on
the theory from the background evolution were presented in \cite{hp5}. For
further applications and extensions of hybrid metric-Palatini theory we refer
the reader in the recent review \cite{hp6}.

To determine exact and analytic cosmological solutions of the fourth-order
hybrid metric-Palatini theory, we consider a mathematical treatment for the
field equations mainly known in Analytic Mechanics. More precisely we apply
the theory of invariant transformations such that to derive invariant surfaces
on the phase space of the dynamical system which describe conservation laws
for the gravitational field equations. That approach has been applied in other
cosmological models and provided many interesting results, see the review in
\cite{ns1}. In this work we investigate the existence of dynamical/contact
symmetries for the gravitational field equations. This kind of symmetry is
more general than the point symmetries and can provide new results on the
study of integrability of dynamical systems \cite{ns2}. The structure of the
paper is as follows.

In Section \ref{sec2}, we present the cosmological model of our consideration,
which is that of hybrid metric-Palatini gravitational theory where in the
Einstein-Hilbert Action Integral a function $f~$ of the $\mathcal{R}$ is the
Palatini curvature scalar is introduced. With the use of Lagrange multiplier
the theory can be written in the equivalent form of a scalar-tensor theory. In
addition, for the cosmological background space we consider that of spatially
flat Friedmann--Lema\^{\i}tre--Robertson--Walker. One interesting property of
that consideration is that the gravitational field equations are described by
a point-like Lagrangian. That is an important characteristic for the rest of
our analysis. The point-like Lagrangian is described by an unknown function
related by the $f\left(  \mathcal{R}\right)  $ function. In Section
\ref{sec3}, we present the basic theory of dynamical symmetries and of
singularity analysis. We perform the complete classification scheme of the
gravitational point-like Lagrangian by determining all the functional forms of
the unknown function for which the gravitational field equations admit
conservation laws given by dynamical symmetries known as contact symmetries,
which provide quadratic in the momentum conservation laws. In Section
\ref{sec4}, for the integrable gravitational systems which followed from the
classification scheme, we write the exact and analytic solutions by reducing
the system in quadratures or by using the singularity analysis such that to
write the analytic solutions with the use of Laurent expansions. The
qualitative behaviour of the exact solutions is also discussed. Finally in
Section \ref{sec5} we draw our conclusions.

\section{Hybrid gravity}

\label{sec2}

Hybrid metric-Palatini gravity is a modified theory of gravity which belongs
to the so-called $f-$theories, where in the Einstein-Hilbert Action Integral a
function $f=f\left(  \mathcal{R}\right)  $ is introduced where $\mathcal{R}$
is the Palatini curvature scalar which is constructed by an independent
connection $\tilde{\Gamma}~$\cite{hp6}.

More precisely, the gravitational Action Integral in hybrid metric-Palatini
theory is defined as \cite{oo1,Capoz}
\begin{equation}
S=\int d^{4}x\sqrt{-g}\left(  R+f(\mathcal{R})\right)  +S_{m},\label{s.01}%
\end{equation}
where $R$ is the metric Ricci curvature scalar and $S_{m}$ describes the
Action Integral for the matter source. Variation with respect to the metric
tensor of (\ref{s.01}) provides the gravitational field equations which are
\cite{Capoz}%
\begin{equation}
G_{\mu\nu}+f^{\prime}(\mathcal{R})\mathcal{R}_{\mu\nu}-\frac{1}{2}%
f(\mathcal{R})g_{\mu\nu}=T_{\mu\nu},\label{s.02}%
\end{equation}
where $G_{\mu\nu}$ is the Einstein tensor for the metric $g_{\mu n}$,
$\mathcal{R}_{\mu\nu}$ is a Ricci tensor constructed by the conformally
related metric $h_{\mu\nu}=f^{\prime}(\mathcal{R})g_{\mu\nu}$, that is,
\cite{Capoz}%
\begin{equation}
\mathcal{R}_{\mu\nu}=R_{\mu\nu}+\frac{3}{2}\frac{f(\mathcal{R})_{,\mu
}f(\mathcal{R})_{,\nu}}{\left(  f(\mathcal{R})\right)  ^{2}}-\frac
{1}{f(\mathcal{R})}f(\mathcal{R})_{;\mu\nu}-\frac{1}{2f(\mathcal{R})}g_{\mu
\nu}\left(  f(\mathcal{R})_{;\kappa\lambda}g^{\kappa\lambda}\right)
\text{,}\label{s.03}%
\end{equation}
prime denotes total derivative with respect to the curvature $\mathcal{R},$
that is $f^{\prime}(\mathcal{R})=\frac{d}{d\mathcal{R}}f\left(  \mathcal{R}%
\right)  $, and $T_{\mu\nu}$ is the energy momentum tensor related with the
Action Integral $S_{m}.$

From (\ref{s.03}) we observe that the field equations (\ref{s.02}) are of
higher-order and thus we can always define a Lagrange multiplier in
(\ref{s.01}) to reduce the order of the field equations and increase the
dimension of the dynamical system, that is, the number of the dependent variables.

Consider the Lagrange multiplier $\lambda$, where the Action Integral becomes
\cite{Capoz}
\begin{equation}
S=\int d^{4}x\sqrt{-g}\left(  R+f(\hat{R})+\lambda\left(  \mathcal{R}-\hat
{R}\right)  \right)  +S_{m}, \label{s.04}%
\end{equation}
while the new variable $\hat{R}$ is defined as $\hat{R}\equiv\mathcal{R}$.

Variation with respect the variable $\hat{R}$ in (\ref{s.04}) gives the
constraint for the Lagrange multiplier $\lambda=f^{\prime}\left(  \hat
{R}\right)  $ such that the Action Integral (\ref{s.04}) is simplified as
follows
\begin{equation}
S=\int d^{4}x\sqrt{-g}\left(  R+f(\hat{R})+f^{\prime}(\hat{R})\left(
\mathcal{R}-\hat{R}\right)  \right)  +S_{m}, \label{s.05}%
\end{equation}
by replacing in $\mathcal{R}=R+\frac{3}{2}\left(  \left(  \frac{f^{\prime
}\left(  \mathcal{R}\right)  }{f\left(  \mathcal{R}\right)  }\right)
^{2}-2\left(  f(\mathcal{R})_{;\kappa\lambda}g^{\kappa\lambda}\right)
\mathcal{\ }\right)  $ and define the new field $\phi$ such that
$\phi=f^{\prime}(\mathcal{R})$ and the function $V\left(  \phi\right)
=\mathcal{R}f^{\prime}\left(  \mathcal{R}\right)  -f\left(  \mathcal{R}%
\right)  $~the gravitational Action Integral can be written as the following
Scalar-tensor theory%
\begin{equation}
S=\int d^{4}x\sqrt{-g}\left(  (1+\phi)R+\frac{3}{2\phi}\partial^{\mu}%
\phi\partial_{\mu}\phi-V(\phi)\right)  +S_{m}. \label{s.07}%
\end{equation}

\subsection{Hybrid cosmology}

In the case of spatially flat Friedmann--Lema\^{\i}tre--Robertson--Walker
(FLRW) spacetime \cite{hp4}
\begin{equation}
ds^{2}=-N^{2}\left(  t\right)  dt^{2}+a^{2}\left(  t\right)  \left(
dx^{2}+dy^{2}+dz^{2}\right)  .\label{s.08}%
\end{equation}
the gravitation field equations (\ref{s.02}) are written in the form of the
Scalar tensor equivalence (\ref{s.07}) as follows~\cite{hp4}%
\begin{equation}
6a\dot{a}^{2}(1+\phi)+6a^{2}\dot{a}\dot{\phi}+\frac{3}{2\phi}a^{3}\dot{\phi
}^{2}-a^{3}V(\phi)=0\label{s.10}%
\end{equation}%
\begin{equation}
4\left(  1+\phi\right)  \dot{H}+4H\dot{\phi}+6\left(  1+\phi\right)
H^{2}-\frac{3}{2}\frac{\dot{\phi}^{2}}{\phi}+2\ddot{\phi}-V\left(
\phi\right)  =0\label{s.11}%
\end{equation}
where $H=\frac{\dot{a}}{a}$ is the Hubble function and we have set $N\left(
t\right)  =1$; the Klein-Gordon equation for the scalar field is%
\begin{equation}
3\frac{\ddot{\phi}}{\phi}-\frac{3}{2}\left(  \frac{\dot{\phi}}{\phi}\right)
^{2}+9H\frac{\dot{\phi}}{\phi}+6\dot{H}+12H^{2}-V_{,\phi}=0.\label{s.12}%
\end{equation}
The latter equation is nothing else that the definition of the scalar
$\mathcal{R}$.

The gravitational field equations (\ref{s.10})-(\ref{s.12}) can be derived by
the singular point-like Lagrangian \cite{hp4}%
\begin{equation}
\mathcal{L}\left(  N,a,\dot{a},\phi,\dot{\phi}\right)  =\frac{1}{N}\left(
6a\dot{a}^{2}(1+\phi)+6a^{2}\dot{a}\dot{\phi}+\frac{3}{2\phi}a^{3}\dot{\phi
}^{2}\right)  +Na^{3}V(\phi), \label{s.13}%
\end{equation}
where the constraint equation is derived from the variation of $\mathcal{L}%
\left(  N,a,\dot{a},\phi,\dot{\phi}\right)  $ with respect to the singular
variable $N$, that is $\frac{\partial L}{\partial N}=0$. On the other hand,
the second-order differential equations are derived with respect to the scale
factor $a\left(  t\right)  $ and the scalar field $\phi\left(  t\right)  $, respectively.

At this point it is important to mention that the field equations
(\ref{s.10})-(\ref{s.12}) constitute a Hamiltonian system, where \ the
constraint equation (\ref{s.10}) is a conservation law for the field equations
with a constraint value. However, because the dimension of the dynamical
system is two, in order to infer about the integrability of the dynamical
system and study the existence of closed-form solutions we should investigate
the existence of additional conservation laws for specific functional forms of
the potential $V\left(  \phi\right)  $, that is, the form of $f(\mathcal{R})$.
In a previous study \cite{hp4} the above system has been constrained according
to the existence of point symmetries.

In this work we investigate the case for which conservation laws quadratic in
the momentum exist for the gravitational field equations. These conservation
laws are related with the existence of dynamical symmetries; the latter are
constructed by the admitted Killing symmetries of the line element which
defines the kinetic energy of the point-like Lagrangian (\ref{s.13}). This
specific line element is also known as minisuperspace.

\section{Dynamical symmetries}

\label{sec3}

We assume the system of second-order differential equations
\begin{equation}
\ddot{q}^{a}=\omega^{a}(t,q,\dot{q}). \label{FL.0}%
\end{equation}
and the infinitesimal transformation with generator%
\begin{equation}
X=\xi(t,q,\dot{q})\partial_{t}+\eta^{a}(t,q,\dot{q})\partial_{q^{a}}
\label{fl.01}%
\end{equation}
which is defined in the jet space $J^{1}\left\{  t,q^{a},\dot{q}\right\}  $.
The vector field $X$ is called a Lie symmetry for the system of differential
equations (\ref{FL.0}) when the transformation under the vector $X$ preserves
the form of the equation and transforms solutions of the dynamical system into
solutions. In the case for which $\frac{\partial\xi}{\partial\dot{q}}%
=\frac{\partial\eta^{\alpha}}{\partial\dot{q}}=0$, the symmetry vector is
called a Lie point symmetry, while when functions $\xi$ and $\eta^{\alpha}$
are linear in $\dot{q}$, the vector field $X$ is called contact symmetry or
dynamical symmetry.

The mathematical condition for the vector field $X$ to be a symmetry vector of
(\ref{FL.0}) is \cite{Stephani,Bluman}%
\begin{equation}
\lbrack X^{\left[  1\right]  },\Gamma]=\lambda(t,q,\dot{q})\Gamma.
\label{fl.02}%
\end{equation}
where $\Gamma$ is the Hamiltonian vector $\Gamma=\frac{d}{dt}=\frac{\partial
}{\partial t}+\dot{q}^{a}\frac{\partial}{\partial q^{a}}+\omega^{a}%
\frac{\partial}{\partial\dot{q}^{a}}$, and $X^{[1]}$ is the first extension of
$X$ defined as $X^{[1]}=\xi(t,q,\dot{q})\partial_{t}+\eta^{a}(t,q,\dot
{q})\partial_{q^{a}}+\left(  \dot{\eta}^{a}-\dot{q}^{a}\dot{\xi}\right)
\partial_{\dot{q}^{a}}$. It is important to mention that in the case of
dynamical symmetries one has an extra degree of freedom which is removed if
one demands an extra condition\thinspace, usually someone requires the gauge
condition $\xi=0$ so that the generator is simplified to $X^{[1]}=\eta
^{a}(t,q,\dot{q})\partial_{q^{a}}+\dot{\eta}^{a}\partial_{\dot{q}^{a}}.$

If (\ref{FL.0}) follows from the variation of the Action Integral $S=\int
L\left(  t,q,\dot{q}\right)  dt$, then under the transformation with generator
$X$, the Euler-Lagrange equations of $L\left(  t,q,\dot{q}\right)  $ are
invariant when there exists a function $f=f\left(  t,q,\dot{q}\right)  $ such
that \cite{Sarlet,SarletCantrijn 81}
\begin{equation}
X^{[1]}L+\frac{d\xi}{dt}L=\frac{df}{dt}. \label{fl.03}%
\end{equation}
Function $f$ is a boundary term introduced to allow for the infinitesimal
changes in Action Integral produced by the infinitesimal change in the
boundary of the domain caused by the transformation of the variables in the
Action Integral \cite{leachnoe1}.

Latter condition is known as the first Noether's theorem \cite{noe1}. In
addition, if condition (\ref{fl.03}) according to the second Noether's theorem
the quantity%
\begin{equation}
\Phi\left(  t,q,\dot{q}\right)  =\xi\left(  \dot{q}^{a}\frac{\partial
L}{\partial\dot{q}^{a}}-L\right)  -\eta^{a}\frac{\partial L}{\partial\dot
{q}^{a}}+f \label{fl.04}%
\end{equation}
is a conservation law for the dynamical system (\ref{FL.0}), that is
$\Gamma\left(  \Phi\left(  t,q,\dot{q}\right)  \right)  \equiv0$. \qquad

For point-like Lagrangians of the form%

\begin{equation}
L\left(  q^{k},\dot{q}^{k}\right)  =\frac{1}{2}\gamma_{ab}\left(
q^{k}\right)  \dot{q}^{a}\dot{q}^{b}-V\left(  q^{k}\right)  , \label{LB.06}%
\end{equation}
and for the generator $X=K_{b}^{a}\left(  t,q^{k}\right)  \dot{q}^{b}%
\partial_{a},$ the Noether symmetry condition (\ref{fl.03}) is simplified as
\cite{Kalotas}
\begin{equation}
K_{\left(  ab;c\right)  }=0~ \label{LB.07}%
\end{equation}%
\begin{equation}
K_{ab,t}=0~~,~f_{,t}=0 \label{LB.08}%
\end{equation}%
\begin{equation}
K_{ab}V^{,b}+f_{,a}=0. \label{LB.09}%
\end{equation}
where $";"$ denotes covariant derivative with respect to the connection
coefficients of the tensor $\gamma_{ab}$. From the symmetry condition
(\ref{LB.08}) we refer that $K_{ab}=K_{ab}\left(  q^{k}\right)  $ and
$f=f\left(  q^{k}\right)  $. Condition (\ref{LB.07}) is a geometric condition
which means that $K_{ab}\left(  q^{k}\right)  $ is a Killing tensor of second
rank for the metric tensor $\gamma_{ab}$, while the existence condition of
function $f\left(  q^{k}\right)  $ is (\ref{LB.09}). Thus, if (\ref{LB.07}%
)-(\ref{LB.09}) are true for a specific potential, the dynamical system with
Lagrangian (\ref{LB.06}) admits the conservation law quadratic in the momentum%
\begin{equation}
\Phi=K_{ab}\left(  q^{k}\right)  \dot{q}^{a}\dot{q}^{b}-f. \label{LB.10}%
\end{equation}
For an recent discussion on quadratic conservation laws in Analytic Mechanics
we refer the reader in \cite{lkarp}.

We continue with the application of the symmetry conditions for the
gravitational point-like Lagrangian of hybrid metric-Palatini cosmology.

\subsection{Integrable dynamical systems in hybrid cosmology}

The cosmological dynamical system described by the point-like Lagrangian
(\ref{s.13}) and for $N\left(  t\right)  =a\left(  t\right)  $, for the
following functional forms of the potential $V\left(  \phi\right)  $,%
\begin{equation}
V_{A}\left(  \phi\right)  =V_{0}\left(  1+\phi\right)  ^{2}~, \label{LB.11}%
\end{equation}%
\begin{equation}
V_{B}\left(  \phi\right)  =V_{0}+V_{1}\phi^{2}~, \label{LB.12}%
\end{equation}%
\begin{equation}
V_{C}\left(  \phi\right)  =~V_{0}\sqrt{\phi}\left(  1+\phi\right)
+V_{1}\left(  1+6\phi+\phi^{2}\right)  ,~ \label{LB.13}%
\end{equation}%
\begin{equation}
V_{D}\left(  \phi\right)  =V_{0}\left(  1+12\phi+16\phi^{2}\right)
+V_{1}\left(  5+20\phi+16\phi^{2}\right)  \sqrt{\frac{\phi}{1+\phi}}~,
\label{LB.14}%
\end{equation}
admits conservation laws quadratic in the momentum. From the potentials we can
always determine the functional form of $f\left(  \mathcal{R}\right)  $ by
solving the Clairaut equation%
\[
V\left(  f^{\prime}\left(  \mathcal{R}\right)  \right)  =\mathcal{R}f^{\prime
}\left(  \mathcal{R}\right)  -f\left(  \mathcal{R}\right)  .
\]

For potential $V_{A}\left(  \phi\right)  $ we have $f_{A}\left(
\mathcal{R}\right)  =-\mathcal{R}+\frac{1}{4V_{0}}\mathcal{R}^{2}$, while from
$V_{B}\left(  \phi\right)  $ we find $f_{B}\left(  \mathcal{R}\right)
=\frac{1}{4V_{1}}\mathcal{R}^{2}-V_{0}$. In a similar way for the remainder of
the potential functions we find the closed form functions $f_{C}\left(
\mathcal{R}\right)  =-\frac{1}{3}\mathcal{R}+\frac{2}{27V_{0}^{2}}%
\mathcal{R}^{3}+\left(  \mathcal{R}^{2}-3V_{0}^{2}\right)  ^{\frac{2}{2}}$ for
$V_{1}=0$, or $f_{C}\left(  \mathcal{R}\right)  =-3\mathcal{R}+\frac{1}%
{4V_{1}}\mathcal{R}^{2}+8V_{1}$ for $V_{0}=0$; $~f_{D}\left(  \mathcal{R}%
\right)  =$ $-\frac{3}{8}\mathcal{R}+\frac{1}{64V_{0}}\mathcal{R}^{2}+\frac
{5}{4}V_{0}$, for $V_{1}=0$. At this point it is interesting to mention that
most of the potentials are related with the $f\left(  \mathcal{R}\right)
=\alpha\mathcal{R}+\beta\mathcal{R}^{2}+\gamma$ for specific values of the
free parameters $\alpha,\beta~$\ and~$\gamma.$

The extra corresponding conservation laws of the gravitational field equations
are calculated for each potential by using Noether's second theorem, they are%
\begin{equation}
\Phi_{A}=\frac{a^{4}}{\phi}\dot{\phi}^{2}~, \label{LB.16}%
\end{equation}%
\begin{equation}
\Phi_{B}=6\dot{a}^{2}-V_{0}a^{4}~, \label{LB.17}%
\end{equation}%
\begin{equation}
\Phi_{C}=-\frac{18}{\sqrt{\phi}}\left(  2\phi\dot{a}^{2}+a\dot{a}\dot{\phi
}\right)  +\frac{3}{4}a^{4}\left(  V_{0}\left(  1+6\phi+\phi^{2}\right)
+16V_{1}\sqrt{\phi}\left(  1+\phi\right)  \right)  ~, \label{LB.18}%
\end{equation}
and
\begin{equation}
\Phi_{D}=\frac{a^{2}}{\sqrt{\phi}}\dot{a}\dot{\phi}-\left(  \frac{2}{3}%
V_{0}\sqrt{\phi}\left(  1+2\phi\right)  +\frac{V_{1}}{6}a^{5}\frac
{1+8\phi\left(  1+\phi\right)  }{\sqrt{1+\phi}}\right)  ~. \label{LB.19}%
\end{equation}

\subsection{Singularity analysis}

Until now we have used Lie's theory and specifically Noether's theorem in
order to study the existence of conservation laws for the gravitational field
equations. Thus in the following section singularity analysis is applied in
order to write the solution of cosmological equations by using Laurent
expansions, thus the basic elements of the singularity analysis are discussed.

Nowadays, the application of singularity analysis is described by the ARS
algorithm \cite{Abl1,Abl2,Abl3}. The first step of the ARS algorithm is based
upon the determination of the leading-order term to prove that a moveable
singularity exists. The determination of the resonances which indicates the
position of the constants of integration is the second step, while the
consistency test is the third and final step of the ARS algorithm. For the
consistency test, we write a Painlev\'{e} Series with exponent and step as
determined in the previous step and study if it is a solution of the original
differential equation. Finally, in the consistency test, the constants of
integration are determined while the coefficients of the Laurent expansions
are derived, which provide the exact form of the algebraic solution for the
given differential equation. In the review article \cite{buntis} various
applications of the ARS algorithm are presented, while the various criteria
which should be satisfied are discussed in detail. Singularity analysis has
been applied before in various cosmological models for the determination of
exact solutions, for instance see \cite{ap1,ap2,ap3,ap4} and references therein.

In the following Section we continue with the determination of exact and
analytic solutions for these specific scalar-tensor potentials.

\section{Exact and analytic cosmological solutions}

\label{sec4}

In order to determine exact and analytical solutions for the gravitational
field equations we make use of the existence of the extra conservation law for
the specific forms of the potential $V\left(  \phi\right)  $ that we found in
the previous Section. The existence of two conservation laws which are
independent and in involution, that is, the dynamical system which is formed
by the gravitational theory for these specific forms of the potentials,
satisfies the criteria of Liouville integrability.

\subsection{Potential $V_{A}\left(  \phi\right)  $}

For the theory with scalar field potential $V_{A}\left(  \phi\right)  $ we
perform the change of variables
\begin{equation}
a=r\cos\theta~,~\phi=\tan^{2}\theta, \label{mm.01}%
\end{equation}
where the point-like Lagrangian is simplified as
\begin{equation}
L\left(  r,\dot{r},\theta,\dot{\theta}\right)  =6\left(  \dot{r}^{2}+r^{2}%
\dot{\theta}^{2}\right)  +V_{0}r^{4}, \label{mm.02}%
\end{equation}
where the gravitational field equations are%
\begin{equation}
\ddot{r}-r\dot{\theta}^{2}-\frac{V_{0}}{3}r^{3}=0~,~\ddot{\theta}+2\frac
{\dot{r}}{r}\dot{\theta}=0 \label{mm.03}%
\end{equation}
with constraint $6\left(  \dot{r}^{2}+r^{2}\dot{\theta}^{2}\right)
-V_{0}r^{4}=0.$ The conservation law becomes $\Phi_{A}=4r^{4}\dot{\theta}^{2}$
from which we find the reduced first-order ordinary differential equation~%
\begin{equation}
\dot{r}^{2}+\frac{\Phi_{A}}{4r^{2}}-\frac{V_{0}}{6}r^{4}=0. \label{mm.04a}%
\end{equation}
The latter equation can be integrated by quadratures as follows
\begin{equation}
\int\frac{dr}{\sqrt{\frac{V_{0}}{6}r^{4}-\frac{\Phi_{A}}{4r^{2}}}}=t-t_{0}.
\label{mm.04}%
\end{equation}

In order to understand this solution we integrate (\ref{mm.04a}) by using
Laurent expansions. Specifically we apply the theory of singularity analysis.
The differential equation (\ref{mm.04a}) admits two leading-order terms with
singular solution%
\begin{equation}
r_{A}\left(  t\right)  =r_{0}\left(  t-t_{0}\right)  ^{-1}~\text{\ with }%
V_{0}=\frac{6}{r_{0}^{6}}%
\end{equation}
and%
\begin{equation}
r_{B}\left(  t\right)  =r_{0}\left(  t-t_{0}\right)  ^{\frac{1}{2}}\text{ with
}\Phi_{A}=-r_{0}^{4}.
\end{equation}
The singular behaviour $r_{A}\left(  t\right)  $ describes a universe in which
the scalar field potential dominates, while $r_{B}\left(  t\right)  $
describes a universe in which the kinematic quantities dominate. Variable
$\theta\left(  t\right)  $ at the leading-order terms becomes%
\begin{equation}
\theta_{A}\left(  t\right)  \simeq\left(  t-t_{0}\right)  ^{3}~\text{and
}\theta_{B}\left(  t\right)  \simeq\ln\left(  \left(  t-t_{0}\right)  \right)
+\theta_{1}%
\end{equation}
from which we infer that near the singularity, $t\rightarrow t_{0}$ the scale
factor is~$a_{A}\left(  t\right)  \simeq r_{A}\left(  t\right)  $ and
$a_{B}\left(  t\right)  \simeq r_{B}\left(  t\right)  $.

We investigate whether equation (\ref{mm.04a}) possesses the Painlev\'{e}
property. To do that we replace $r\left(  t\right)  =r_{0}\left(
t-t_{0}\right)  ^{p}+m\left(  t-t_{0}\right)  ^{p+s}\,,$ where $p=-1,\frac
{1}{2}$ in (\ref{mm.04a}) and we linearize around the value $m\simeq0$. The
leading-order terms of the leading-order terms vanish when $s=-1$, which
indicates that the singularities are movable poles for both solutions and we
can write the analytic solution by using Laurent expansions.

For $p=-1$ we find the solution%
\begin{equation}
r_{A}\left(  t\right)  =r_{0}\left(  t-t_{0}\right)  ^{-1}+r_{6}\left(
t-t_{0}\right)  ^{5}+r_{12}\left(  t-t_{0}\right)  ^{11}+r_{18}\left(
t-t_{0}\right)  ^{17}+...~\text{with}~V_{0}=\frac{6}{r_{0}^{6}}~,
\end{equation}
where $r_{6}=\frac{\Phi_{A}}{56r_{0}^{3}},~r_{12}=-\frac{9\Phi^{2}}%
{81536r_{0}^{7}},...$ .

On the other hand for $p=\frac{1}{2}$ we find
\begin{equation}
r_{B}\left(  t\right)  =r_{0}\left(  t-t_{0}\right)  ^{\frac{1}{2}}%
+r_{3}\left(  t-t_{0}\right)  ^{\frac{1}{2}+3}+r_{6}\left(  t-t_{0}\right)
^{\frac{1}{2}+6}+...~\text{with }\Phi_{A}=-r_{0}^{4}~
\end{equation}
and $r_{3}=\frac{_{r_{0}^{3}V_{0}}}{24}~,~r_{6}=\frac{9r_{3}^{2}}{14r_{0}%
},...$ .

We remark that the spacetime near the singularity solutions becomes%
\begin{equation}
ds^{2}=\left(  t-t_{0}\right)  ^{2p}\left(  -dt^{2}+dx^{2}+dy^{2}%
+dz^{2}\right)  ~,~p=-1,\frac{1}{2}.
\end{equation}
For $p=-1,$ we have the de Sitter solution, while for $p=\frac{1}{2}$ the
stiff fluid solution follows.

\subsection{Potential $V_{B}\left(  \phi\right)  $}

For potential $V_{B}\left(  \phi\right)  $ we apply the change of variables
$\phi=\frac{\psi^{2}}{a^{2}}$, where the point-like Lagrangian is written in
the simple form%
\begin{equation}
L\left(  a,\dot{a},\psi,\dot{\psi}\right)  =6\left(  \dot{a}^{2}+\dot{\psi
}^{2}\right)  +V_{0}a^{4}+V_{1}\psi^{4}.
\end{equation}
The gravitational field equations are%
\begin{equation}
\ddot{a}-\frac{V_{0}}{3}a^{3}=0~,~\ddot{\psi}-\frac{V_{1}}{3}\psi^{3}=0
\end{equation}
with constraint equation $6\left(  \dot{a}^{2}+\dot{\psi}^{2}\right)
-V_{0}a^{4}-V_{1}\psi^{4}=0$. Easily it follows that the gravitational field
equations are equivalent with the system%
\begin{equation}
6\dot{a}^{2}-V_{0}a^{4}=h_{1}~,~6\dot{\psi}^{2}-V_{1}\psi^{4}=-h_{1},
\end{equation}
which means that the analytic solution is expressed in terms of the Jacobi
elliptic function. When $h_{1}=0$, exact solution of the field equations
is~$a\left(  t\right)  =\frac{24}{V_{0}}\left(  t-t_{0}\right)  ^{-2}$ and
$\psi\left(  t\right)  =\frac{24}{V_{1}}\left(  t-t_{0}\right)  ^{-2}$.
Moreover, with the use of the singularity analysis, by following the steps for
potential $V_{A}\left(  \phi\right)  $, the analytic solution can be expressed
in terms of Laurent expansion as follows%

\begin{equation}
a\left(  t\right)  =\frac{24}{V_{0}}\left(  t-t_{0}\right)  ^{-2}+a_{6}\left(
t-t_{0}\right)  ^{4}+a_{12}\left(  t-t_{0}\right)  ^{10}+..., \label{mm.02a}%
\end{equation}
with $a_{6}=-\frac{V_{0}h_{1}}{4032}~,...~$. Similarly, for the field
$\psi\left(  t\right)  $ we find
\begin{equation}
\psi\left(  t\right)  =\frac{24}{V_{0}}\left(  t-t_{0}\right)  ^{-2}+\psi
_{6}\left(  t-t_{0}\right)  ^{4}+\psi_{12}\left(  t-t_{0}\right)  ^{10}+...~.
\label{mm.02b}%
\end{equation}

However, we have not considered the case for which $V_{0}=0$, which means that
the cosmological constant term is zero. In that consideration the exact
solution for the scale factor is $a\left(  t\right)  =h_{1}\left(
t-t_{0}\right)  .$ The exact solution with $V_{0}$ and $h_{1}=0$ provides that
the its geometry is described by the line element%
\begin{equation}
ds^{2}=\left(  t-t_{0}\right)  ^{4}\left(  -dt^{2}+dx^{2}+dy^{2}%
+dz^{2}\right)  .
\end{equation}

The latter spacetime describes a universe dominated by a dust fluid source.

\subsection{Potential $V_{C}\left(  \phi\right)  $}

We apply the coordinate transformation%
\begin{equation}
a=u+v~,~\phi=\left(  \frac{u-v}{u+v}\right)  ^{2},
\end{equation}
where the point-like Lagrangian is simplified as
\[
L\left(  u,\dot{u},v,\dot{v}\right)  =6\left(  \dot{u}^{2}+\dot{v}^{2}\right)
+\bar{V}_{0}u^{4}+\bar{V}_{1}v^{4}
\]
with $\bar{V}_{0}=V_{0}+4V_{1}$ and $\bar{V}_{1}=4V_{1}-V_{0}$. Thus, in a
similar way as above the gravitational field equations are reduced to the
system%
\begin{equation}
6\dot{u}^{2}-\bar{V}_{0}u^{4}=h_{1}~,~6\dot{v}^{2}-V_{1}v^{4}=-h_{1},
\end{equation}
while in terms of Laurent expansion the analytic solution is expressed as in
(\ref{mm.02b}), in which the leading-order behaviour describes the matter
dominated era.

\subsection{Potential $V_{D}\left(  \phi\right)  $}

In order to write the solution of the gravitational field equations for the
potential $V_{D}\left(  \phi\right)  $ we prefer to work in the parabolic
coordinates given by the coordinates%
\begin{equation}
a=uv~,~\phi=\left(  \frac{u^{2}-v^{2}}{2uv}\right)  ^{2},
\end{equation}
in which the point-like Lagrangian becomes%
\begin{equation}
L\left(  u,\dot{u},v,\dot{v}\right)  =6\left(  u^{2}+v^{2}\right)  \left(
\dot{u}^{2}+\dot{v}^{2}\right)  +\frac{\left(  V_{0}+V_{1}\right)
u^{10}+\left(  V_{0}-V_{1}\right)  v^{10}}{u^{2}+v^{2}}.
\end{equation}

The constraint equation can be written with the use of the momentum
$p_{u}=\frac{\partial L}{\partial\dot{u}},~p_{v}=\frac{\partial L}%
{\partial\dot{v}}$ as follows%
\begin{equation}
\frac{1}{6\left(  \dot{u}^{2}+\dot{v}^{2}\right)  }\left(  p_{u}^{2}+p_{v}%
^{2}-6\left(  V_{0}+V_{1}\right)  u^{10}+6\left(  V_{0}-V_{1}\right)
v^{10}\right)  =0.
\end{equation}
Hence the Hamilton-Jacobi equation is%
\begin{equation}
\frac{1}{6\left(  \dot{u}^{2}+\dot{v}^{2}\right)  }\left(  \left(
\frac{\partial}{\partial u}S\left(  u,v\right)  \right)  ^{2}+\left(
\frac{\partial}{\partial v}S\left(  u,v\right)  \right)  ^{2}-6\left(
V_{0}+V_{1}\right)  u^{10}-6\left(  V_{0}-V_{1}\right)  v^{10}\right)  =0
\end{equation}
with $p_{u}=\frac{\partial S}{\partial u}$ and $p_{v}=\frac{\partial
S}{\partial v}$, that is,%
\begin{equation}
S\left(  u,v\right)  =-\int\sqrt{h_{1}+6\left(  V_{0}+V_{1}\right)  u^{10}%
}du-\int\sqrt{-h_{1}+6\left(  V_{0}-V_{1}\right)  v^{10}}dv.
\end{equation}

With the use of the latter Action the gravitational field equations are
reduced to the following system of first-order ode%
\begin{equation}
12\left(  u^{2}+v^{2}\right)  \dot{u}=-\sqrt{h_{1}+6\left(  V_{0}%
+V_{1}\right)  u^{10}}%
\end{equation}
and%
\begin{equation}
12\left(  u^{2}+v^{2}\right)  \dot{v}=-\sqrt{-h_{1}+6\left(  V_{0}%
-V_{1}\right)  v^{10}}.
\end{equation}

From there we find the parametric solution%
\begin{equation}
\frac{du}{dv}=\frac{\sqrt{h_{1}+6\left(  V_{0}+V_{1}\right)  u^{10}}}%
{\sqrt{-h_{1}+6\left(  V_{0}-V_{1}\right)  v^{10}}}.
\end{equation}

In the special case for which $h_{1}$ we find the special solution%
\begin{equation}
\frac{u^{-4}}{\sqrt{V_{0}+V_{1}}}=\frac{v^{-4}}{\sqrt{V_{0}-V_{1}}}\text{.}%
\end{equation}

From the later expression it follows that $u^{2}\simeq v^{2}$, from which it
follows $\dot{u}=-c_{0}u^{3},$that is,~$u\left(  t\right)  =\left(  2c\left(
t-t_{0}\right)  \right)  ^{-\frac{1}{2}}$, from which it follows that
$a\left(  t\right)  \simeq\left(  t-t_{0}\right)  ^{-1}$, which describes the
de Sitter universe.

\section{Conclusion}

\label{sec5}

We applied the theory of Lie symmetries for the determination of conservation
laws for the cosmological field equations in hybrid metric-Palatini theory. In
particular, we made use the symmetry conditions for the existence of quadratic
in the momentum conservation laws, such that to constrain the unknown
functional form of the metric-Palatini theory. For simplicity on our
calculations we worked in the equivalent scalar-tensor theory, from where we
found that four potential functions for which the gravitational field
equations admit at least a second conservation law and the cosmological field
equations are Liouville integrable.

For the four different scalar-tensor potentials, we were able to reduce the
field equations into a system of two first-order ordinary differential
equations, which can be solved in all cases in quadratures. For specific
values of the initial conditions we were able to write closed-form exact
solutions; while we made use of the singularity analysis such that to write
the analytic solutions with Laurent expansions. We found that important
cosmological eras in the cosmological solutions are provided by the analytic
solutions of our models.

This work contributes in the subject of integrability of modified
gravitational theories. The existence of integrable models for a given
gravitational model is essential for the mathematical validity of a theory.
Indeed, the existence of integrable trajectories is important such that to
relate numerical solutions with actual/real solutions of the dynamical system.
Moreover, the integrable models can be used as toy models for the study of the
specific theory. In a forthcoming work we plan to investigate further these
analytic solutions and study the asymptotic behaviour and the resulting
cosmological history.

\end{document}